# BULGE/DISK SEGREGATION IN THE UNIVERSE

Gary A. Mamon

DAEC, Observatoire de Paris-Meudon, F 92195 Meudon, FRANCE

# BULGE/DISK SEGREGATION
# IN THE UNIVERSE

Gary A. Mamon

DAEC, Observatoire de Paris-Meudon, F 92195 Meudon, FRANCE

**Abstract.** The observations of bulge/disk segregation in the Universe are reviewed with a focus on whether the observed segregation in clusters is local or global, and whether there is bulge-disk segregation on large-scales. The high concentration of bulge-rich galaxies in the cores of clusters of galaxies can be accounted for by several popular physical processes: 1) biased early elliptical formation, 2) abortion of disk formation by tidal destruction of the gas reservoirs that fuel such disks, and 3) ram pressure stripping of the gas in disks by intergalactic gas, 4) merging of spiral galaxies into ellipticals (which is the main focus of this review). A global scenario is outlined incorporating each of these processes.

## 1. Introduction

Part of the beauty in galaxies resides in their varied morphologies. In his pioneering galaxy classification study, Hubble (1926) defined galaxy types as a function of 1) bulge to disk ratio, 2) openness of the spiral arms, 3) resolution of bright knots (HII regions) in the spiral arms. At one end lay the early-type gas-poor galaxies, called *ellipticals*, whose characteristics are smooth elliptical isophotes, with surface brightness falling off roughly as $R^{-2}$ (the Hubble law first introduced by Reynolds 1913). Since then, ellipticals have been also described by the $R^{1/4}$ law (de Vaucouleurs' 1948), where surface brightness falls off as $\exp(-cR^{1/4})$. At the other extreme, lay the late-type *spiral* galaxies, whose spiral arms lay within in a flattened disk, later shown to have a 3D density profile falling off exponentially both along the plane (de Vaucouleurs 1959) and perpendicular to it. Spiral galaxies of lesser late types exhibit small bulges, that resemble elliptical galaxies. In between, ellipticals and spirals lie the *lenticular* galaxies, also called S0 galaxies, whose bulges are nearly as luminous as the disks, but whose disks are devoid of gas and spiral arms. Hubble & Humason (1931) already noted that ellipticals tended to lie in dense regions while spirals were more isolated galaxies. This was first quantified by Dressler (1980), see §2. The present review is devoted to the dynamical explanations of the variations of bulge to disk ratio in varying environments. There will be no mention here of barred spirals and little focus on star formation. The reader is encouraged to read the recent excellent reviews on the origins of the Hubble sequence by Larson (1993) and Evrard (1993), as well on the details of galaxy mergers by Barnes & Hernquist (1992).

# 2. Observed Morphological Segregation

## 2.1. Is morphological segregation local or global?

In his pioneering study of morphological segregation in 55 clusters, Dressler (1980) studied the variations of the fraction of spirals, lenticulars and ellipticals as a function of both projected number density and radius (distance from the cluster center). He concluded that these morphological abundance variations were stronger for the local criterion (projected local number density), whereas he saw little radial segregation of morphologies. This local view of morphological segregation, was enhanced by Postman & Geller (1984), who devised a deprojection algorithm and applied it to both the CfA galaxy catalog (Huchra *et al.* 1983), and Dressler's sample of 55 clusters. Indeed Postman & Geller found a *universal* morphology-density relation applying to all regimes of galaxy clustering, from very loose groups to rich clusters.

This picture of *local* morphological segregation would remain unchallenged until the experiments of Sanromà & Salvador-Solé (1990), who took Dressler's data, and placed each galaxy in a given cluster at a new position with the same clustocentric radius, but with a random polar angle. The morphology-density relation obtained from these scrambled artificial clusters turned out to be virtually identical to that of the original clusters. Hence, the morphologies of galaxies seem to depend more on their distance to their parent cluster, a *global* characteristic, than to local inhomogeneities. Soon thereafter, Whitmore & Gilmore (1991) reanalyzed Dressler's (1980) original data, being careful to split the wide (half-Mpc) bin of smallest clustocentric radii into smaller radial bins. They show that, with decreasing radius, the spiral fraction steadily declines, the fraction of lenticulars increases up to 100 kpc, then declines dramatically, while the fraction of ellipticals is nearly constant within 200 kpc, then drastically rises to reach a level of $\simeq 55\%$ in the cluster center. In their sequel, Whitmore, Gilmore & Jones (1993) show that galaxy morphologies depend more on clustocentric radii than on local number density, thus confirming Sanromà & Salvador-Solé's global view of morphological segregation, although there remains a local effect in clusters and also in the field (Charlton, Whitmore & Gilmore 1993). They also demonstrate that these trends of morphological abundance are robust when one changes the definition of the cluster center, from the barycenter to the position of the central luminous D or cD galaxy, to the center of the X-ray isophotes.

Now, it may be that Dressler's cluster sample consists mainly of smooth regular clusters, so that the local morphological segregation that may come out of the minority of irregular ones is washed out. Indeed, about one-third of well-studied clusters exhibit large-scale substructure (Jones & Forman 1992), and one might then expect these irregular clusters to show less radial morphological segregation as local morphological segregation. Unfortunately, there are no studies known to this reviewer that have considered the morphological segregation in irregular clusters.

Returning to Dressler's original work, one notices in his cluster diagrams, that galaxies of the same morphology tend to aggregate together: *i.e.,* there are clumps that are rich in lenticulars, other clumps rich in ellipticals. The same effect has been noticed by Andreon (1993), who showed through extensive Monte-Carlo simulations that such positional segregation is statistically significant in comparison with a random Poisson distribution of galaxies. If anything, this finding brings back to focus the idea of local morphological segregation. However, the distribution of lenticulars must be compared with Monte-Carlo simulations where the morphologies fractions versus radius are drawn from Whitmore *et al.*'s (1993) radial distribution, and since these authors find that the fraction of lenticular galaxies peaks at a few hundred kpc from the cluster center, the clumping of lenticulars may arise from azimuthally irregular distributions within Whitmore *et al.*'s radial distribution.

### 2.2. Is there large-scale bulge-disk segregation?

Because elliptical galaxies are restricted to the cores of clusters of galaxies, one sees a large-scale bulge-disk segregation. However, is there any such segregation left when one peels out clusters from the large-scale galaxy distribution? This question was first answered by Davis & Geller (1976), who showed that the angular correlation functions in the CfA galaxy catalog (Huchra *et al.* 1983) were considerably steeper for ellipticals than for spirals, even after removing the objects sitting in the Virgo and Coma clusters. This large-scale morphological segregation has been confirmed by Santiago & Strauss (1992), who investigated the 3D galaxy distributions in the CfA catalog. However, these authors suggest that, once cluster galaxies are removed, the Local Supercluster is more sharply defined by the *spirals*, which seems to contradict the idea that ellipticals are more clustered on large scales. It would seem useful to repeat these analyses on larger samples and with a variable peeling off of dense regions. In any event, the presence of large-scale morphological segregation would favor morphological segregation processes at galaxy formation.

## 3. Theories of Disk/Elliptical Segregation

Various mechanisms have been proposed to set the morphological type at galaxy formation or alter the morphological type of a galaxy during its subsequent evolution, and often with the consequence of explaining the observed bulge-disk segregation. An incomplete list would include: gas removal by collisions (Spitzer & Baade 1951), IGM ram pressure (Gunn & Gott 1972), interstellar gas evaporation by conduction from the hot intergalactic medium (Cowie & Songaila 1977), viscous turbulent stripping (Nulsen 1982) and gas consumption by bursts of star formation (Larson, Tinsley & Caldwell 1980), all turning spirals into lenticulars; spiral bars heated by accretion evolving into enhanced bulges (Pfenniger & Norman 1990); disk heating by accretion (Quinn & Goodman 1986); abortion of disks by tidal disruption of protodisks (Larson *et al.* 1980; Dressler 1980); disks merging into ellipticals (Toomre 1977); bulges built by mergers (Schweizer 1992).

### 3.1. Morphological segregation set at galaxy formation

Cosmologists like to see the characteristics of astrophysical objects set by cosmological initial conditions. In particular, there has been plenty of work to understand galaxies, or at least their halos, from the primordial density fluctuation spectrum. Indeed, at early epochs (for example at recombination around $z = 1000$), the Universe was nearly homogeneous. All theories of the growth of structures by gravitational instability have been based upon the assumption that at these early epochs, there was sufficient inhomogeneity to drive gravitational instability to generate the extreme density contrasts witnessed today, say between a galaxy and the manyfold more tenuous intergalactic medium. It has thus been a great relief to have the COBE satellite unambiguously detect these fluctuations, on scales of 10° (Smoot *et al.* 1992).

In this cosmological framework, present-day elliptical galaxies have often been identified with the end products of the highest small-scale fluctuations in a Cold Dark Matter (hereafter CDM) primordial density fluctuation spectrum (starting with Blumenthal *et al.* 1984). The physical reasoning is that the highest peaks in the primordial density fluctuation spectrum collapse the fastest, and therefore have no time to be spun-up by the gain of angular momentum through tidal torques with their neighboring environment. However, two sets of numerical simulations (Barnes & Efstathiou 1987; Zurek *et al.* 1988) have shown that the angular momentum normalized to that of a particle in circular orbit of the same energy $[\lambda = J/J_{\rm circ}(E)]$ is independent of local density.

Without delving into the problems with spin angular momentum, one can simply ask, as Evrard, Silk & Szalay (1990) whether the peaks that are above $3\sigma$ at early epochs lead to present-day elliptical galaxies, further identifying lenticular galaxies with fluctuations between 2.5 and $3\sigma$ and spiral galaxies with fluctuations between 2 and $2.5\sigma$. Because clusters of galaxies would, in such a context, naturally represent the high peaks of the primordial density fluctuation spectrum on large scales, it would then be easier to reach the high fluctuations required for ellipticals if these formed in clusters. A similar line of reasoning has been advocated by West (1993) to explain the observed (Harris & Racine 1979) high frequency of globular clusters (high peaks on very small scales) in ellipticals. From analytical considerations, Evrard *et al.* produce the correct abundance ratios between ellipticals, lenticulars and spirals, but are not able to check the normalization as a function of total galaxy density. They then perform cosmological simulations that nicely reproduce Dressler's observed fraction of present-day ellipticals as a function of projected galaxy number density, however the dividing line between lenticulars and spirals is not explained. Despite the beauty of this model, it seems to this reviewer, that the physical processes involved in galaxy formation (with cooling *and* heating) are too complex to parametrize in a simple over density at early epochs.

One can also revert to hydrodynamics to understand the present-day morphologies of galaxies. In a pioneering study, Gott & Thuan (1976) argued that if the gas collapses before it forming stars one obtains a gas rich system, probably resembling a present-day spiral galaxy, and conversely in the denser systems the stars form before the system collapses and one gets an elliptical. Thus Gott & Thuan already appreciated that ellipticals would arise form high peaks in the primordial density field, and also stated that such peaks were more likely to occur in clusters. Reasoning from a more dissipational point of view, Larson (1976) then explained how one can get star formation before collapse in the denser systems: simply by collisions of gas clouds which are all the more frequent that the system is dense and has high velocity dispersion. Of course star formation is a two step operation: first the gas has to cool and then it has to form stars. The cooling time constraints were incorporated within a cosmological context with Blumenthal *et al.*'s (1984) analysis with CDM primordial density fluctuations. Unfortunately, considerations of cooling times are not sufficient to understand galaxy formation. Indeed, as shown by Blanchard, Valls-Gabaud & Mamon (1992), the naïve application of such physics leads to a an *overcooling* phenomenon in which one would expect nearly all the gas in the Universe to have converted into stars by today, in contrast with the observations, and one has then to resort to a *heating* period at the epoch of galaxy formation $z \simeq 3$ to regulate the cooling.

### 3.2. Aborted disks

The beauty of the aborted disk scenario is that it provides an explanation for the origin of the hot IGM gas in clusters (Whitmore, Gilmore & Jones 1993), which is simply made of the gas reservoirs that are stripped by the global cluster potential and thus prevented to cool and dissipate into galactic disks. This scheme is thus intimately connected to galaxy formation.

A test of this idea would be the absence of intergalactic gas in groups and poor clusters where the global potential tides are too weak to strip these gas reservoirs. In other words, the fraction of matter in gas should be higher in rich clusters than in poor clusters and groups. In the unified scheme for groups (Mamon, in these proceedings), loose groups are still collapsing today and have not seen much interpenetration of their dark matter halos, and one thus does not expect any disk abortion. The few cases of detection of hot IGM gas in groups leads to similar gas fractions as in rich clusters (see Henriksen & Mamon 1993), but one of these groups is compact and probably dense in 3D, whereas the others may just be dense binaries or triplets. Note that, in the aborted disk scenario, some spiral galaxies are still present in dense regions (for example NGC 2276 in the NGC 2300 loose group), simply because these spirals form in relatively isolated regions and then fall into the denser regions (Shaya & Tully 1984). These spirals must interact with their environment and thus appear different from isolated spirals. One check has been to compare the gradient of the rotation curves of spirals galaxies in different environments, and whereas preliminary measurements indicated that the spirals sitting in the centers of clusters had *decreasing* rotation curves (Whitmore, Forbes & Rubin 1988), this fact seems to have been contradicted with more accurate measurements of a similar sample of galaxies (Amram *et al.* 1993).

### 3.3. Ram pressure stripping

Although ram pressure stripping has been demonstrated to produce the correct morphological segregation, for any choice of cluster properties (Solanes & Salvador-Solé 1992), it is not clear how the stripping of the gas will affect the bulges and disks. Dressler (1980) first reasoned that the greater B/D ratio of lenticulars reflects the fact that S0s have brighter bulges than spirals, which thus argues against ram pressure stripping as responsible for producing S0s. However, Solanes, Salvador-Solé & Sanromà (1989) showed that if one considered galaxy samples that are bulge-magnitude limited instead of total-magnitude limited, the data seems to favor the opposite result that the greater B/D ratio of lenticulars reflects the fact that their disks are fainter while their bulges are comparable to those of spirals. This revives the possibility of ram pressure stripping being an efficient B/D converter. It remains to be understood why S0s are not observed to be fainter than spirals. Note that a while back, Kent (1981) had argued that if disks are faded by some process like tidal disruption of protodisks, one would reproduces Dressler's (1980) morphological segregation.

# 4. Realistic Galaxy Formation through Simulations

A recent set of CDM cosmological simulations with gas dynamics, by Steinmetz & Müller (1993), sheds much light on the formation of bulges and disks. These simulations start with a mixture of dark matter particles and a gaseous component (modeled with the SPH smoothed particle hydrodynamics technique, see Friedli in these proceedings). The gas can convert into stars when the gas density and temperature are sufficiently high and low, respectively. The important new ingredient in these simulations is the feedback of energy and metals to the gas (by supernova explosions). When the authors simulate the formation of an isolated spiral galaxy, they are able to reproduce remarkably well, not only a bulge+disk morphology, but also the kinematical and chemical properties of both components, with in addition to the thin disk of recently formed (less than 2 Gyr old) stars, a thick disk of older stars. Indeed, Steinmetz & Müller find that the stars that make up the inner bulge are old and metal-rich and form from gas originating from the highest peaks of the primordial density fluctuation spectrum, whereas the outer bulge (often called *spheroid* or *halo*) is composed of old metal-poor stars. Star formation occurs in two phases, first the bulge stars and later the disk stars (mainly in one burst).

An important aspect of this work is that the final bulge to disk ratio is a function of the amount of power on small scales in the primordial density fluctuation spectrum (the more small-scale power the more important is the bulge). Steinmetz & Müller are unable to produce elliptical galaxies with these simulations of single galaxy formation. However, they have run similar simulations where the initial conditions lead to two galaxies merging on a parabolic orbit at redshift $z \simeq 2$, and their final product at $z = 1$ resembles very much a present-day elliptical, except that the metallicity gradient produced is greater than is observed. The interesting thing to note here is that the ellipticals do not form from the merger of two spirals, but instead from the merger of two *proto-spirals* who are in the midst of their collapse and star formation phase. Steinmetz & Müller conclude that both ellipticals and the bulges of spirals originate from local maxima in the primordial density fluctuation spectrum, although in addition a merger of two proto-spirals is required to produce ellipticals.

# 5. A Merger Model for Groups and Clusters

Galaxy mergers remain one of the principal candidates for altering the bulge to disk ratios of galaxies. A serious difficulty assessed early on (Ostriker 1980) is that elliptical galaxies tend to lie in dense regions with large velocity dispersions, whereas mergers require slow encounters, and hence are only effective in regions of low velocity dispersions. One way around this argument is to suppose that ellipticals form by mergers in low velocity dispersion groups that later fall into clusters. In what follows, is described an alternative approach (Mamon 1992) in which mergers occur among the slowest encounters within rich clusters.

### 5.1. Merging in isolated spherical clusters

Consider first an isolated spherical cluster. The inner parts of this cluster have decoupled from their initial Hubble expansion at an early epoch, and later, subsequent shells of matter fall into the cluster, and quickly virialize as they rebound, settling to a clustocentric distance of roughly half their initial turnaround radius. These shells carry a few relatively isolated galaxies, each possessing huge halos of dark matter (a few hundred kpc in size). How does the merging rate evolve within this shell? In the first stages of its initial Hubble expansion, merging may occur because the densities are high, but the time available for merging is short. Near turnaround, the densities are low and merging is infrequent. When the shell falls into the cluster, there is a first phase where it is moving too fast relative to the cluster, so that encounters between its galaxies and the cluster's are too rapid to lead to merging. Once the shell rebounds out of the cluster center, the huge galaxy halos are tidally stripped by the global cluster potential, the remaining galaxies (with small halos, tens of kpc in size) virialize with the rest of the cluster, spending most of their time at their orbital apocenter, again roughly half their initial turnaround radius. This is the time when merging can start to be effective.

How effective can mergers be in a region of high velocity dispersion? Because the galaxy orbits are similar to those of test particles within the potential of the dominant dark matter, one can compute the merger rate by integrating the merger cross-section over a Maxwellian distribution of relative velocities. The merger cross-section has been estimated a while ago in the first numerical experiments of galaxy-galaxy collisions (Roos & Norman 1979; Gerhard 1981). The resultant merger rate $k$ can be written as

$$nk = n\langle \sigma v \rangle = \text{Cst}\, n r_g^2 v_g K(v_{\text{cl}}/v_g)\,, \tag{1}$$

where $\sigma$ is the merger cross-section (and depends on the relative encounter velocity $v$), $r_g$ and $v_g$ are the test galaxy's half-mass radius and internal velocity dispersion, respectively, and $K$ is the effective merger rate (for given number density of galaxies) and is a function of the ratio of cluster velocity dispersion $v_{\text{cl}}$ to galaxy internal velocity dispersion.

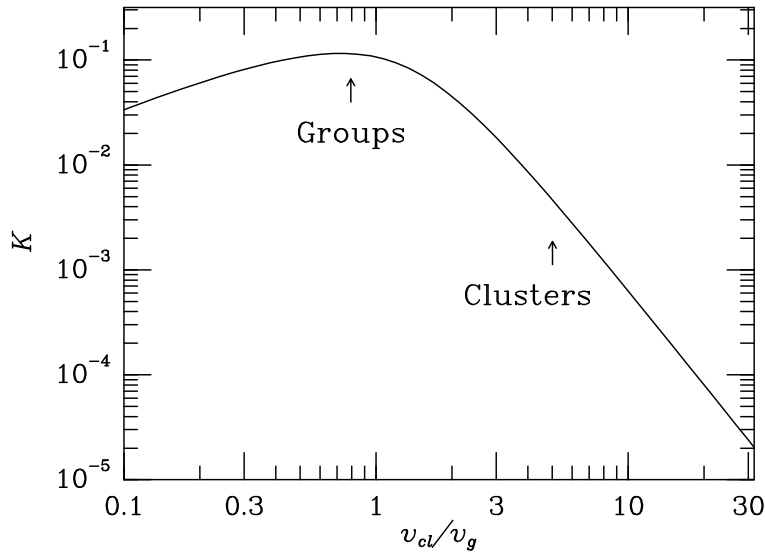

**Figure 1.** Effective merger rate $K$ (see eq. [1]) versus system velocity dispersion (in units of galaxy internal velocity dispersion)

Figure 1 shows the dimensionless merger rate $K$. As the cluster velocity dispersion dispersions increase, the merger rate decreases because a decreasing fraction of collisions are slow enough, while on the contrary, in the limit of small cluster velocity dispersions, there are simply not enough collisions for mergers. The optimum case for merging is when the velocity dispersion dispersion of the galaxy system is similar to the internal velocity dispersions of galaxies, as is the case in groups of galaxies. *For the typical velocity dispersions of rich clusters, the merger rate is not zero, but simply 25 times less effective than in groups.*

Writing (to first-order) the fraction of galaxies that are the products of mergers as $f = nK$, assuming here that time is not a factor, *i.e.*, that galaxies in different environments have similar amounts of time available for merging, one concludes that in typical (loose) groups, which are roughly 100 times less dense than the cores of rich clusters, the fraction of merger products should be roughly 4 times smaller than in the cluster cores. If one identifies merger products with elliptical morphologies, then Postman & Geller's (1984) universal morphology-density relation is at odds with the merger model, which predicts a factor 25 enhancement of the elliptical fraction in low velocity dispersion groups of same galaxy number density as high velocity dispersion clusters. The discrepancy may be caused by the fact that their largest density bin for loose groups contains the Virgo cluster (Whitmore *et al.* 1993), as well as the fact that loose groups are not virialized and their member galaxies are collapsing together for the first time, hence the lack of mergers (see Mamon in these proceedings).

On the other hand, Hickson's (1982) compact groups obey a morphology-density relation which is offset relative to Postman & Geller's, as they are considerably more spiral-rich than the cluster cores of similar density (Mamon 1986). This can readily be explained if compact groups are chance alignments within larger loose groups (Mamon 1986). The morphological fraction in a chance alignment should mimic that of its parent group. Moreover, because the frequency of chance alignments is very dependent on the ratio of the apparent radius of the compact group to the radius of the surrounding loose group (Walke & Mamon 1989), the densities of chance alignments will typically be some fraction of that of their surrounding loose groups. Hence, the morphology-density relation is expected to parallel that of loose groups.

### 5.2. Local morphological segregation

The trends of elliptical fraction versus number density and clustocentric radius can be reproduced by the simple merger model. For the high velocity dispersions of clusters, the effective merger rate scales as $K \sim v_{\rm cl}^{-3}$ (see Fig. 1). It is then easy to show that the merger rate and resultant elliptical fraction is

$$f \sim nk = {\rm Cst}\, nG^2 m^2/v_{\rm cl}^3 \,. \qquad (2)$$

Now in the spherical cosmological infall model one has $\rho \sim R^{-9/4}$, $M \sim R^{3/4}$, and $v_{\rm cl} \sim R^{-1/8}$, for the local mass density, enclosed mass and local velocity dispersion, respectively. Now the mass of a galaxy will be set by the tides from the global cluster potential as it passes through the cluster core the first time. One can show that the galaxy radius is then limited to a size proportional to its orbital pericenter. For an $\Omega = 1$ cosmology, all galaxy orbits are self-similar, so that orbit apocenter is proportional to orbit pericenter. Hence, the galaxy mass is limited to $m \sim M(R) \sim R^{3/4}$. The galaxy number density varies as $n \sim \rho/m \sim R^{-3}$ yielding the fraction of merger products scaling as

$$f \sim R^{-9/8} \sim n^{3/8} . \qquad (3)$$

To the next order of precision the reaction $D + D \to E$ (where $D$ and $E$ stand for disks and ellipticals, respectively), produces a fraction of merger products set by the equation $df/dt = nk(1 - f)$ with the solution

$$\frac{f - f_i}{(1 - f)(1 - f_i)} = \int_{t_i}^{t_0} nk \, dt \simeq nk(t_0 - t_i) ,$$

where $f_i$ is the fraction of merger products before the shells rebound within clusters, and where the local merger rate, after rebound, is taken as time-independent, as is expected in the pile-up approximation to spherical cosmological infall (Gott 1975), where successive shells settle at successively greater radii, hence *all local properties are fixed in time and the merger rate too.*

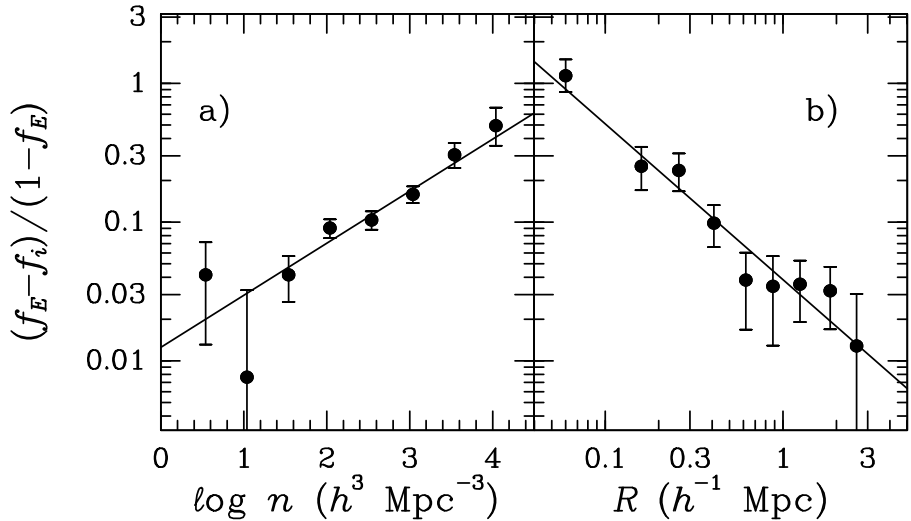

**Figure 2.** *a)* Morphology-density relation for ellipticals ($f_E$), with best fit field elliptical fraction ($f_i$) of 7.3%. *Points*: Observed relation (Postman and Geller 1984). *Lines:* Power-law predictions from equation (3). *b)* Morphology-radius relation for ellipticals ($f_E$), with best fit field elliptical fraction of 13.0%. *Points:* Observed relation (Whitmore *et al.* 1993) restricted to R < 3 h$^{-1}$ Mpc. *Lines:* Power-law predictions from equation (3).

Figure 2 show the fits to the morphology-density and morphology-radius relations of Postman & Geller (1984) and Whitmore, Gilmore & Jones (1993), respectively. *The merger model reproduces extremely well the trends of morphology with local density and* *clustocentric radius*. The normalization of these fits depend on the assumed galaxy (mean) mass $m = 2 \times 10^{11} M_\odot$, which corresponds to $M/L = 60\,h$, which is a reasonable value for galaxies with halos of a few tens of kpc.

### 5.3. Morphology-velocity dispersion relations

From equation (2), one may expect the elliptical fraction of clusters to decrease with cluster temperature as $f \sim v_{\rm cl}^{-3}$. Edge & Stewart (1991) show that the spiral fraction in clusters decreases with increasing cluster temperature, seemingly at odds with the prediction above. However they also show that hot clusters are rich, with $n \sim T^{0.9}$ (and also confirm $T \sim v_{\rm cl}^2$). If the mean galaxy mass is uncorrelated with cluster temperature, then the merger model is fails to reproduce the correlations observed by Edge & Stewart. Unfortunately, nothing is known on how mean galaxy mass correlates with cluster temperature. But Edge & Stewart (1991) state that the luminosity of the brightest cluster member varies as $L_1 \sim T$. Thus, if the mean galaxy mass scales as the luminosity of the brightest cluster member then the merger model does predict the observed morphology velocity dispersion relation for clusters. Note that the disk abortion model works well for clusters with deep potentials, hence high velocity dispersions, and thus naturally explains the trend of increasing elliptical fraction with increasing cluster velocity dispersion (Whitmore *et al.* 1993).

In compact groups, morphologies are more correlated with velocity dispersion than with number density (Hickson, Kindl & Huchra 1988). Although this trend is heavily contaminated by a strong morphology-distance trend (Tikhonov 1990; Mamon 1990; Whitmore 1990), the morphology-velocity dispersion relation is still statistically significant once the distance correlations are removed (Whitmore 1992; Mendes de Oliveira 1992), and more so than the morphology-density relation. If merger products are identified with ellipticals, then one expects to see a morphology-density relation in compact groups but not a morphology-velocity dispersion relation, since the function $K$ is near maximum for groups (Fig. 1).

This dilemma disappears once one brings back *time* into the picture. This is done through the fundamental surface analysis of groups (Mamon, in these proceedings), which shows that the nature of compact groups depends on their velocity dispersion: the low velocity dispersion compact groups are chance alignments within collapsing loose groups, the intermediate velocity dispersion compact groups are normal groups at full collapse, and the high velocity dispersion compact groups are the groups that have had enough time to collapse and rebound, and thus witness mergers.

### 5.4. Morphology versus time

As mentioned above, the merger model predicts that local merger rate is constant in time. At first glance, this seems at odds with the Butcher-Oemler (1978) effect, which states that relatively distant ($z > 0.2$) clusters harbor a significant greater fraction of blue galaxies. Indeed, recent investigations point to these blue galaxies being interacting galaxies (Lavery & Henry 1988; Dressler 1993). Now if the local merger rate is constant in time, then the rate of galaxy interactions may also be roughly constant in time, and there should be as many galaxies interacting now than at $z > 0.2$.

More precisely, the fraction of galaxies currently undergoing a merging interaction should be proportional to the rate of mergers, hence to $nk \sim r^{-9/8}$. Therefore one should see a higher fraction of interacting galaxies in cluster cores than in cluster envelopes. Also, since the local merger rate is constant in time, *the fraction of merging galaxies within a fixed metric radius should also be constant in time*. However, as time passes, clusters accrete lower density shells, so that *the merger model predicts that the mean fraction of merging galaxies within the entire cluster should decrease with time*, which perhaps explains the Butcher-Oemler effect. Obviously, the observed radial trends of the fraction of interacting galaxies must be measured in both local and distant clusters to confront this merger model prediction.

### 5.5. Hierarchical cluster merging

Clusters of galaxies are rarely isolated and spherical, hence the underlying assumptions of the merger model seem optimistic. There are two extreme limits where cluster-cluster collisions may alter the results of the present model: 1) accretion of groups and small clusters onto large clusters, and 2) mergers of clusters of similar mass.

If a large cluster accretes a group or small cluster of much smaller mass, this will only slightly perturb the morphological mix and segregation within the large one. If, however, clusters *only* grow by the accretion of groups and small clusters, then the final morphological mix and segregation will be much more a function of that in the infalling groups. But if these groups are accreted at all times during the cluster history, then assuming that they settle at some fixed fraction of their turnaround radius (maximum distance to the cluster), as was assumed above for the individual infalling galaxies, then the trends of morphological segregation should be preserved, only the normalization will be altered to lower galaxy $M/L$. Could groups settle at very low radii, by dissipating their orbital energy by dynamical friction (and thus modify the morphological segregation)? Because the dynamical friction time is likely not to be proportional to the free-fall time (see Mamon, in these proceedings for a discussion of dynamical timescales in clusters), the slopes of the morphology-density and morphology-radius relations would be altered. However, calculations of the orbital evolution of infalling groups (González-Casado, Mamon & Salvador-Solé 1993, see also Mamon in these proceedings) show that infalling groups are tidally disrupted at first passage through the cluster core, before dynamical friction has time to operate. Thus, in summary, *the accretion of groups and small clusters into large clusters has little effect on the slopes of morphological segregation*.

If two clusters of similar mass merge together one may first think that the morphological segregation in each may be erased by the merger. Now in numerical simulations of galaxy-galaxy mergers, the particles that are the most bound to each individual galaxy end up as the most bound in the merged galaxy (Barnes 1988). It is most reasonable to assume that the same situation will hold for mergers of galaxy clusters, in which case the morphological segregation may be the same as for the case of isolated clusters. It remains to be checked quantitatively whether the correct trends of morphological segregation would come out of this cluster merger scenario. Especially because there should be some level of erasure of the morphological segregation at each merger. But such cases of mergers between clusters of similar masses will be relatively rare (see Lacey & Cole 1993; Kauffmann & White 1993), so that in between two mergers, the morphological segregation may reestablish itself. Moreover, when two clusters of similar mass merge together it is likely that galaxies from each will also merge together, especially those in the cores of each cluster, where the densities are higher. This effect could balance the low-level erasure of morphological segregation arising from the mixing of the two cluster populations.

### 5.6. Possible model extensions

The merger model can be refined in several ways: 1) The model "chemistry" can be made self-consistent by considering the reactions $D+D \to E$, $D+E \to E$ and $E+E \to E$. The predicted morphological segregation is no longer described by power-laws, but by functions that are close to power-laws. 2) The model can be generalized to multi-masses. This is all the more interesting that the mass ratio of merging galaxies can be used to set the new bulge-disk ratio in terms of the old one. For example, all merging collisions with mass ratios ranging from 1 : 1 to 3 : 1 would lead to ellipticals, 3 : 1 to 10 : 1 to lenticulars (see Charlton *et al.* 1993). Thus, one can attempt to derive the full Hubble sequence with an extension of the current merger model, along the lines as proposed by Schweizer (1992). 3) The cosmology can be generalized to $\Omega < 1$, where again the convenient power-laws disappear. Note that the infall model used to model mergers can be also used to model tidal stripping of the halos to study gradients of the rotation curves, as well as the disk abortion scenario.

### 5.7. A hybrid formation/merging scenario

Whereas the formation mechanisms for morphological segregation have been based upon the idea that galaxy morphologies are set by the initial height of the peak of the primordial density field (see §3.1 above), one may adapt this approach to general morphological segregation, replacing the concept "galaxy" by that of "group" (see also Mamon 1993). The statistics of the primordial density field force dense groups of galaxies to form near larger scale peaks that are the progenitors of rich clusters. These groups turnaround and collapse internally at early epochs. From the discussion in §5.1, galaxy merging is extremely rapid in these dense groups, so that when the dense group falls into its neighboring cluster it is elliptical-rich (or a single luminosity elliptical galaxy if merging has completed before infall into the cluster). A detailed analysis of this model is in preparation.

## 6. Discussion and Summary

The competition among several physical processes to win the title of the responsible for bulge-disk segregation is far from being settled. It may very well be that each of these processes dealt with in this review plays an important role in developing this morphological segregation. The presence of morphological segregation on large scales seems to imply that formation processes are at work, at least in this large-scale, "linear" regime. The height of the peaks in the primordial density field set the timescale for collapse, which can be compared with the timescale for gas cooling and subsequent star formation. With such tools as the Press & Schechter (1974) formalism it is relatively easy to study the statistics of the primordial density field. This can be extended to dense groups, which will form near clusters and where very rapid merging will lead to ellipticals. Disk abortion by tidal stripping of the gas reservoirs of disks has emerged as a strong contender for the details of the bulge-disk segregation, in part because the slow infall of disks makes their formation quite hazardous. Presumably evolutionary processes are important too. Indeed, ram pressure stripping should play a significant role in the formation of lenticulars, while merging may rebuild the entire Hubble sequence. It is unfortunate that rapid collapse and merging both give rise to similar looking elliptical galaxies, and more detailed modeling of both elliptical formation mechanisms must be performed and confronted to accurate observations. For example, which of these two scenarios reproduces better the distribution of apparent ellipticities of elliptical galaxies? To conclude with a positive note, it appears that we are now at a stage where we can build a global quasi-analytical model where both formation and evolutionary processes can be incorporated and hence tested.

*Acknowledgements:* The author is indebted to Dave Burstein, Henry Ferguson, Eduard Salvador-Solé, Simon White and Brad Whitmore for useful discussions.

# References

Amram P., Sullivan W.T., Balkowski C., Marcelin M. & Cayatte V.: 1993, ApJ 403, L59
Andreon S.: 1993, submitted to A&A
Barnes J.: 1988, ApJ 331, 699
Barnes J. & Efstathiou G.: 1987, ApJ 319, 575
Barnes J.E. & Hernquist L.: 1992, ARAA 30, 705
Blanchard A., Valls-Gabaud D. & Mamon G.A.: 1992, A&A 264, 365
Blumenthal G.R., Faber S.M., Primack J.R. & Rees M.J.: 1984, Nature 311, 517
Butcher H. & Oemler A.: 1978, ApJ 219, 18
Charlton J., Whitmore B.C. & Gilmore D.M.: 1993 in *Groups of Galaxies*, ed. O. Richter (San Francisco: A.S.P.) in press
de Vaucouleurs G.: 1948, Annales d'Astrophys 11, 247
de Vaucouleurs G.: 1959, in *Handbuch der Physik* vol 53 (Berlin: Springer Verlag), 275
Dressler A.: 1980, ApJ 236, 351
Dressler A.: 1993, Sky & Tel 85–4, 22
Edge A.C. & Stewart G.C.: 1991, MNRAS 252, 428
Evrard A.E.: 1992, in *The Physics of Nearby Galaxies: Nature or Nurture?*, ed. T.X. Thuan, C. Balkowski & J. Tran Van Thanh (Gif-sur-Yvette: Frontières) 375
Evrard A.E., Silk J. & Szalay A.: 1990, ApJ 265, 13
Forbes D.A. & Whitmore B.C.: 1989, ApJ 339, 657
Gerhard O.E.: 1981, MNRAS 197, 179
Gott J.R.: 1975, ApJ 201, 296
Gott J.R. & Thuan T.X.: 1976, ApJ 204, 649
Gunn J.E. & Gott J.R.: 1972, ApJ 176, 1
Harris W.E. & Racine R.: 1979, ARAA 17, 241
Henriksen M.J. & Mamon G.A.: 1993, submitted to ApJ Letters
Hickson P.: 1982, ApJ 255, 382
Hickson P., Kindl E. & Huchra J.P.: 1988, ApJ 331, 64
Hubble E.: 1926, ApJ 64, 321

Hubble E. & Humason M.L.: 1931, ApJ 74, 43
Huchra J.P, Davis M., Latham D.W. & Tonry J.L.: 1983, ApJS 52, 89
Jones C. & Forman W.: 1992, in *Clusters and Superclusters of Galaxies*, ed. A.C. Fabian (Dordrecht: Kluwer) 49
Kauffmann G. & White S.D.M.: 1993, MNRAS 261, 921
Lacey C.G. & Cole S.: 1993, MNRAS 262, 627
Larson R.B.: 1976, MNRAS 176, 31
Larson R.B.: 1992, in *The Physics of Nearby Galaxies: Nature or Nurture?*, ed. T.X. Thuan, C. Balkowski & J. Tran Van Thanh (Gif-sur-Yvette: Frontières) 487
Larson R.B., Tinsley B.M. & Caldwell C.N.: 1980, ApJ 237, 692
Lavery R.J. & Henry J.P.: 1988, ApJ 330, 596
Mamon G.A.: 1986, ApJ 307, 426
Mamon G.A.: 1990, in *Paired and Interacting Galaxies*, ed. J.W. Sulentic, W.C. Keel & J.M. Telesco (Washington: NASA) 619
Mamon G.A.: 1992, ApJ 401, L3
Mamon G.A.: 1993, in *Groups of Galaxies*, ed. O. Richter (San Francisco: A.S.P.) in press
Mendes de Oliveira C.: 1992, PhD thesis, Univ. of British Columbia
Nulsen P.E., 1982, MNRAS 198, 1007
Ostriker J.P.: 1980, Comm Ap, 8, 177
Pfenniger D. & Norman C.: 1990, ApJ 363, 391
Postman M. & Geller M.J.: 1984, ApJ 281, 95
Quinn P.J. & Goodman J.: 1986, ApJ 309, 472
Reynolds J.H.: 1913, MNRAS 74, 132
Roos N. & Norman C.A.: 1979, A&A 95, 349
Sanromà M. & Salvador-Solé E.: 1990, ApJ 360, 16
Santiago B.X. & Strauss M.A.: 1992, ApJ 387, 9
Schweizer F.: 1992, in *The Physics of Nearby Galaxies: Nature or Nurture?*, ed. T.X. Thuan, C. Balkowski & J. Tran Van Thanh (Gif-sur-Yvette: Frontières) 283
Shaya E.J. & Tully R.B.: 1984, ApJ 281, 56
Simien F. & de Vaucouleurs G.: 1986, ApJ 302, 564
Smoot G.F. *et al.* 1992, ApJ 396, L1
Solanes J.M. & Salvador-Solé E.: 1992, ApJ 395, 91
Solanes J.M., Salvador-Solé E. & Sanromà M.: 1989, AJ 98, 798
Spitzer L. & Baade W.: 1951, ApJ 113, 413
Steinmetz M. & Müller E.: 1993, submitted to A&A Letters
Tikhonov N.A.: 1990, in *Paired and Interacting Galaxies*, ed. J.W. Sulentic, W.C. Keel & J.M. Telesco (Washington: NASA) 105
Walke D.G. & Mamon G.A.: 1989, A&A 295, 291
West M.J.: 1993, MNRAS in press
Whitmore B.C.: 1990 in *Clusters of Galaxies*, ed. W.R. Oegerle, M.J. Fitchett & L. Danly (Cambridge: Cambridge University Press) 139
Whitmore B.C.: 1992, in *The Physics of Nearby Galaxies: Nature or Nurture?*, ed. T.X. Thuan, C. Balkowski & J. Tran Van Thanh (Gif-sur-Yvette: Frontières) 351
Whitmore B.C., Forbes D.A. & Rubin V.C.: 1988, ApJ 333, 542
Whitmore B.C. & Gilmore D.M.: 1991, ApJ 367, 64
Whitmore B.C., Gilmore D.M. & Jones C.: 1993, ApJ 407, 489
Zurek W.H., Quinn P.J. & Salmon J.K.: 1988, ApJ 330, 519